\RequirePackage{fix-cm}
\documentclass[smallextended]{svjour3}       
\smartqed  
\usepackage{graphicx}
\usepackage{mathrsfs}
\usepackage{graphicx}
\usepackage{amssymb}
\usepackage{amsmath}
\usepackage{cases}
\usepackage{caption}
\usepackage{amsfonts,color,morefloats}
\usepackage{hyperref}
\DeclareMathOperator{\ord}{ord}
\newtheorem{Thm}{Theorem}
\newtheorem{Def}[Thm]{Definition}
\newtheorem{Lem}[Thm]{Lemma}
\newtheorem{Cor}[Thm]{Corollary}

\newtheorem{Exp}[Thm]{Example}
\newtheorem{Rem}[Thm]{Remark}
\makeatletter
 \@addtoreset{equation}{section}
 
\makeatother
\journalname{}
\begin{document}

\title{Linear complexity of  generalized cyclotomic sequences of order $4$ over $\mathbb{F}_{l}$}


\author{Yuhua Sun$^{1}$, Yang Yan$^{2}$, Fei Li$^{*,3}$, Tongjiang Yan$^{1}$, Hui Li$^{4}$}


\institute{Y. Sun, T. Yan, \at $1$. College of Science, China University of Petroleum
Qingdao, Shandong 266580, China\\
\email{sunyuhua-1@163.com, yantoji@163.com}\\
Y. Yan, \at $2$. National Engineering Laboratory for Information Security Technologies
,  Institute of Information Engineering,
             Chinese Academy of Sciences,
              Beijing, 100195,  China\\
              \email{yanyang9021@iie.ac.cn}\\
F. Li,  corresponding author\at $3$. School of Statistics and Applied Mathematics,  Anhui University of Finance and Economics,
 Bengbu City,   Anhui Province, 233030, China\\
              \email{cczxlf@163.com }\\
H. Li,  \at $4$. State Key Laboratory of Integrated Service Networks,  Xidian University,
 Xi$^{,}$an,  Shaanxi Province, 710071, China\\
              \email{lihui@mail.xidian.edu.cn }           
}

\date{Received: date / Accepted: date}
\maketitle

\begin{abstract}
 Generalized cyclotomic  sequences of period $pq$ have several desirable randomness properties if the two primes $p$ and $q$ are chosen properly. In particular,   Ding deduced the exact formulas for the autocorrelation and the linear complexity of these sequences of order $2$. In this paper, we consider the generalized sequences of order $4$. Under certain  conditions,  the linear complexity  of these sequences of order $4$ is developed  over a finite field $\mathbb{F}_{l}$. Results show that in many cases they have high linear complexity.\\


\end{abstract}

\section{Introduction}

Let $l$ be a prime number and $\mathbb{F}_{l}$ denote a finite field with $l$ elements. A sequence $s=s_0s_1\ldots s_{n-1}\ldots$  is called to be $n$-periodic if $s_{i}=s_{i+n}$ for all $i\geq0$. Periodic sequences with certain properties are widely used in software testing, radar systems, stream ciphers and so on. For cryptography applications, the linear complexity is an important factor. It is defined to be the length of the shortest linear feedback shift register which generates this sequence. The generalized cyclotomic sequences have been described and studied for the past decades and have resulted in numerous constructions \cite{BLX,BW,Ding97,Ding98,DingH,FG,HYW}. These are  interesting since they have a number of attractive randomness properties \cite{Ding12,MW,WJL152,WJL15,WJLG14}.

Let $p$ and $q$ be two distinct odd primes with $\gcd(p-1, q-1)=d$. Define $n=pq$ and $e=(p-1)(q-1)/d$. The Chinese Reminder Theorem guarantees that there exists a common primitive root $g$ of both $p$ and $q$. Let $x$ be an
integer satisfying
\begin{equation}\label{eq1.1}
x\equiv g\pmod{p},\  x\equiv1\pmod{q}.
\end{equation}
Whiteman proved that \cite{ALW}
$$
\mathbb{Z}_{n}^{*}=\{g^{s}x^{i}:s=0,1,\cdots,e-1;i=0,1,\cdots d-1\},
$$
where $\mathbb{Z}_{n}^{*}$ denotes the set of all invertible elements of the residue class ring $\mathbb{Z}_{n}$.

 The generalized cyclotomic classes $D_{i}$ $(0\leq i\leq d-1)$ of order $d$ with respect to $n$   are defined by \cite{ALW}
\begin{equation}\label{defD}
D_{i}=\{g^{s}x^{i}:s=0,1,\ldots, e-1\},\qquad (0\leq i\leq d-1),
\end{equation}
where the multiplication is that of $\mathbb{Z}_{n}$. Clearly, the cosets $D_{i}$  depend on the choice of the common primitive root $g$ if $d\geq$ 3. It is not hard to prove that \cite{ALW}
$$
\mathbb{Z}_{n}^{*}=\bigcup_{i=0}^{d-1}D_{i},\ D_{i}\cap D_{j}=\emptyset, \ (i\neq j),
$$
where  $\emptyset$ denotes  the empty set. Define
$$
P=\{p,2p, \ldots, (q-1)p\}=p\mathbb{Z}_{q}^{*}, \   Q=\{q,2q,\ldots, (p-1)q\}=q\mathbb{Z}_{p}^{*}.
$$
Then $\{0\}$, $P$, $Q$ and $D_{i}$ $(0\leq i\leq d-1)$ is a partition of $\mathbb{Z}_{n}$. Let $S$ be an nonempty subset of $\{0,1, \cdots, d-1\}$ and
$$
\Sigma_{S}=\bigcup_{i\in S}D_{i}.
$$

We define the following binary sequence $s=s_0s_1s_2\ldots s_{n-1}\ldots$ of period $n$  by
\begin{equation*}
s_i=\left\{\begin{array}{ll}
             1, & \textrm{ if $i \bmod{n}\in \cup P\cup \Sigma_{S}$},\\
             \rho , & \textrm{ if $i\bmod{n}\in \{0\}$}, \\
             0, & \textrm{ otherwise},
           \end{array}
           \right.
\end{equation*}
where  $i\geq0$ and  $\rho\in \{0,1\}$. For $\rho=0$ and $S=\{0\}$,  the linear complexity of these sequences over $\mathbb{F}_2$ have been calculated by Ding \cite{Ding97} with $d=2$ and Hu et al. \cite{HYW}with $d=4$. Furthermore, for $S=\{0,2,\cdots, d-2\}$  and $\rho=0$,  Ding \cite{Ding12} determined the linear complexity of the two-prime sequences over finite field $\mathbb{F}_{l^{m}}$ where $\gcd(l,n)=1$,  and   used these sequences to construct several classes of  cyclic codes over a finite field  with optimal or almost optimal property. In this paper, we only consider the case $\rho=1$, $d=4$ and $S=\{0,1\}$. Under the assumption that $\frac{n-1}{4}\equiv0\pmod{l}$ or $l\notin D_0$, we calculate the linear complexity of these sequences over the finite field $\mathbb{F}_{l}$. Results show that in many cases these sequences have high linear complexity.

This paper is organized as follows. Section $2$ presents basic notations and results of periodic sequences and the  generalized cyclotomy \cite{ALW}. In section $3$, we give an expression  for the linear complexity of the generalized cyclotomic sequences over $\mathbb{F}_{l}$. In the last section, we present concluding remarks of this paper.

\section{Preliminaries}

Firstly, we give the definition and formula of linear complexity of periodic sequences
over  a finite field. See  \cite{CDR} or \cite{LiNi}  for more details.

Let $l$ be a  prime number and let $s=s_0s_1\cdots\ s_{n-1}\cdots$ be a periodic sequence over  $\mathbb{F}_{l}$  with period $n$, where $s_i\,\in\mathbb{F}_{l}$ for $i\geq0$. The sequence $s$ can be viewed as a power series
$$s^{\infty}(x)=\sum\limits_{i=0}^{\infty}s_ix^{i}=\frac{s(x)}{1-x^n},\qquad s(x)=s_0+s_1x+s_2x^2+\cdots+s_{n-1}x^{n-1}\in\mathbb{F}_{l}[x]
$$
in the power series ring $\mathbb{F}_l[[x]]$  .

Let $h(x)=\gcd(s(x),1-x^n)$ , then
$$
s^{\infty}(x)=\frac{w(x)}{v(x)},\qquad v(x)=\frac{1-x^{n}}{h(x)}\qquad w(x)=\frac{s(x)}{h(x)}
$$
where $w(x),v(x),h(x)\in\mathbb{F}_l[x]$.
\begin{Def}
The polynomial $v(x)$ is called the minimal polynomial of the periodic sequence $s$ over $\mathbb{F}_{l}$. The  $\deg\,v(x)= n-\deg\,h(x)$ is called the linear complexity of the sequence $s$ over $\mathbb{F}_{l}$, which is denoted by $L_l(s)$.
\end{Def}

Indeed $L_l(s)$ is the length of the shortest linear feedback shift register which generates the sequence $s$.

\label{}
If $\gcd(n,l)=1$, then $1-x^n$ has $n$ distinct zeros $\zeta_n^i\ (0\leq\,i\leq\,n-1)$ in the algebraic closure $\Omega_l$ of $\mathbb{F}_l$. It is easy to see that
\begin{equation}\label{linearcomp}
L_l(s)=n-\#\{i:0\leq\,i\leq\,n-1,s(\zeta_n^i)=0\}.
\end{equation}

In order to determine the linear complexity of generalized cyclotomic sequences, we introduce  generalized cyclotomy.

 Let the symbols be as in the introduction and $d=\gcd(p-1,q-1)=4$.
The  generalized  cyclotomic numbers  of order $4$ with respect to $n$ is defined by
$$
(i,j)=|(D_{i}+1)\cap D_{j}|
$$
for $0\leq i$, $j\leq3$.

By a well-known theorem (\cite{LWJ}, P. 128), there are exactly two representations of $n$ in the form $n=a^{2}+4b^{2}$ with $a\equiv1\pmod{4}$ and the sign of $b$ indeterminate.

 Let $g_{1}$ and $g_2$  be a  fixed primitive root of $p$ and $q$, respectively.  For $i=1,2$, let $x_i$, $y_{i}$ be integers given uniquely by
 \begin{eqnarray}
 p=x_1^{2}+4y_1^{2},\qquad  q=x_2^{2}+4y_2^{2},\qquad   x_{1}\equiv x_{2}\equiv1\pmod{4}, \nonumber \\
 2y_{1}\equiv -(-g_{1})^{\frac{p-1}{4}}x_{1}\pmod{p}, \qquad 2y_{2}\equiv -(-g_{2})^{\frac{q-1}{4}}x_{2}\pmod{q}.
 \end{eqnarray}
 Define $a$, $b$ to be integers satisfying
 \begin{equation}\label{eq3.17}
 a=x_1x_2+4\left(\frac{2}{p}\right)\left(\frac{2}{q}\right)y_1y_2, \qquad b=x_1y_2-\left(\frac{2}{p}\right)\left(\frac{2}{q}\right)x_2y_1,
 \end{equation}
 where $\left(\frac{\cdot}{\cdot}\right)$ denotes the Legendre symbol.

It is clear that $a\equiv1\pmod{4}$ and $n=a^{2}+4b^{2}$ is one of the two representations of $n$. The following lemma shows that the generalized cyclotomic numbers of order $4$ with respect to $n$ depend uniquely on this representation.
\begin{table}[ht]
\centering
\caption{$p\not\equiv q\pmod{8}$}\label{tal:cyclotomy40}
\begin{tabular}{lllll}
\hline
$(i,j)$ \qquad& 0 \qquad& 1 \qquad& 2 \qquad& 3   \\
\hline
0 \qquad&   A \qquad& B \qquad& C \qquad& D \\
1 \qquad& E \qquad& E \qquad& D \qquad& B \\
2  \qquad& A \qquad& E \qquad& A  \qquad& E \\
3 \qquad& E \qquad& D \qquad& B  \qquad& E \\
\hline
\end{tabular}
\end{table}

\begin{table}[ht]
\centering
\caption{$p\equiv q\pmod{8}$}\label{tal}
\begin{tabular}{lllll}
\hline
$(i,j)$ \qquad& 0 \qquad& 1 \qquad& 2 \qquad& 3  \\
\hline
0 \qquad&   A \qquad& B \qquad& C \qquad& D\\
1 \qquad& B \qquad& D \qquad& E \qquad& E \\
2  \qquad& C \qquad& E \qquad& C   \qquad& E\\
3 \qquad& D \qquad& E \qquad& E   \qquad& B \\
\hline
\end{tabular}
\end{table}

\begin{Lem}[\cite{FG}, Theorem IV.1.]\label{lem2.2}
Let $p\equiv q\equiv1\pmod{4}$  be two distinct primes with the fixed primitive roots $g_1$ and $g_2$, respectively. $M=\frac{(p-2)(q-2)-1}{4}$, and $a$, $b$ are the integers defined in \eqref{eq3.17}.

If $p\not\equiv q\pmod{8}$, then  in    $\autoref{tal:cyclotomy40}$  $8A=-a+2M+3$, $8B=-a-4b+2M-1$, $8C=3a+2M-1$, $8D=-a+4b+2M-1$, $8E=a+2M+1$.

If  $p\equiv q\pmod{8}$, then in $\autoref{tal}$  $8A=3a+2M+5$, $8B=-a+4b+2M+1$, $8C=-a+2M+1$, $8D=-a-4b+2M+1$, $8E=a+2M-1$.
\end{Lem}

\section{Generalized cyclotomic sequences of order $4$}
Throughout this section, let $p$ and $q$ be two distinct odd primes with $\gcd(p-1,q-1)=4$. Define $n=pq$ and $e=(p-1)(q-1)/4$. Let $l$ be a prime  and satisfy $\gcd(l,n)=1$.

 The generalized cyclotomic sequence $s$ of order $4$ of period $n$ is defined by
\begin{equation}\label{df3.1}
s_{i}=\left\{\begin{array}{ll}
    1, & \textrm{ if $i\bmod{n}\in \{0\}\cup P\cup D_0\cup D_1$},  \\
    0, & \textrm{ otherwise},
  \end{array}
  \right.
\end{equation}
where $D_0$, $D_1$ are defined by \eqref{defD} and $P=p\mathbb{Z}_{q}^{*}$. Here in this paper, we treat it as a sequence over a finite field $\mathbb{F}_l$, where $\gcd(l,n)=1$.

 Denote $\ord_{n}(l)$ the multiplicative order of $l$ modulo $n$. Let $\zeta_{n}$  be an $n$-th primitive root of unity over $\mathbb{F}_{l^{\ord_{n}(l)}}$. For the sequence $s$ defined by \eqref{df3.1}, we know
$$
s(x)=1+\sum_{i\in P}x^{i}+\sum_{i\in D_0}x^{i}+\sum_{i\in D_1}x^{i},
$$
and
\begin{equation}\label{eq3.16}
s(1)=1+\frac{(p+1)(q-1)}{2}\bmod{l}.
\end{equation}
Define $\delta$ as follows
\begin{equation}\label{eq3.1.1}
\delta=\left\{\begin{array}{ll}
                1, & \textrm{ if $l \mid 1+ \frac{(p+1)(q-1)}{2}$},  \\
                0, & \textrm{ otherwise}.
              \end{array}
              \right.
\end{equation}

Note that the generalized cyclotomic classes of order $2$ are given by
\begin{eqnarray*}
C_0=D_0\cup D_2,\qquad C_1=D_1\cup D_3.
\end{eqnarray*}
Define $\eta_0=\sum_{i\in C_0}\zeta_{n}^{i}$. The following lemma  has been proven in \cite{Ding12}.
\begin{Lem}[\cite{Ding12}, Lemma 3.13] \label{lem3.1}
If $n\equiv1\pmod{4}$, then we have
$$
\eta_0(1-\eta_0)=-\frac{n-1}{4}.
$$
\end{Lem}
Hence, $\eta_0\in \{0,1\}$ if and only if $(n-1)/4\equiv0\pmod{l}$.

To compute the linear complexity of $s$, we need to compute $\gcd(x^{n}-1, s(x))$. For this purpose, we require a number of auxiliary results.
\begin{Lem}[\cite{Ding97}, Lemma 5]\label{lem2.5}
Let $m$ be the least common multiple of two positive integers $m_1$ and $m_2$. The system of congruences
\begin{eqnarray}\label{eq3.4}
x\equiv a_1\pmod{m_1},\qquad x\equiv a_2\pmod{m_2}
\end{eqnarray}
has solutions if and only if
\begin{equation}\label{eq3.5}
\gcd(m_1,m_2)| a_1-a_2,
\end{equation}
where $a|b$ means that $a$ divides $b$. When the condition \eqref{eq3.5} holds, the system of the congruences of \eqref{eq3.4} has only one solution modulo $m$.
\end{Lem}
\begin{Lem}\label{lem2.4}
For $a\in D_{j}$, then $aD_{i}=D_{i+j \bmod{4}}$.
\end{Lem}
\begin{proof}
To prove this lemma, we need to prove $x^{4}\in D_0$  for  the integer $x$ defined by \eqref{eq1.1}.

By the generalized Chinese Reminder Theorem, there exists an integer $s$  with $0\leq s\leq e-1$ such that
\begin{eqnarray*}
\left\{\begin{array}{l}
         s\equiv 4\pmod{p-1}, \\
         s\equiv 0\pmod{q-1}.
       \end{array}
       \right.
\end{eqnarray*}
That is, the integer $s$ satisfies
$$
\left\{\begin{array}{l}
         x^{4}\equiv g^{s}\pmod{p}, \\
          x^{4}\equiv g^{s}\pmod{q}.
       \end{array}
       \right.
$$
This $s$ is unique.
Hence, $x^{4}\equiv g^{s}\pmod{pq}$ and $x^{4}\in D_0$.
\end{proof}

Since $\zeta_{n}$ is an $n$-th primitive root of unity over $\mathbb{F}_{l^{\ord_{n}(l)}}$, we have
\begin{align}\label{eq3.1}
\sum_{i\in P}\zeta_{n}^{i}=-1,\qquad  \sum_{i\in Q}\zeta_{n}^{i}=-1.
\end{align}
By the definition of $\zeta_{n}$, we have
\begin{align*}
\zeta_{n}^{n}-1&=(\zeta_{n}-1)\left(1+\sum_{i\in P}\zeta_{n}^{i}+\sum_{i\in Q}\zeta_{n}^{i}+\sum_{j=0}^{3}\sum_{i\in D_{j}}\zeta_{n}^{i}\right)\\
&=0.
\end{align*}
Together with \eqref{eq3.1}, we get
\begin{equation}\label{eq3.3}
\sum_{j=0}^{3}\sum_{i\in D_{j}}\zeta_{n}^{i}=1.
\end{equation}

Define $t(x)=\sum_{i\in D_1}x^{i}+\sum_{i\in D_2}x^{i}$.
\begin{Lem}\label{lem2.3}
Let the symbols be the same as before. Then
$$
s(\zeta_{n}^{a})=\left\{\begin{array}{ll}
                         s(\zeta_{n}),  &  a\in D_0, \\
                          t(\zeta_{n}), &  a\in D_1, \\
                          -\left(s(\zeta_{n})-1\right), &  a\in D_2 ,\\
                          -\left(t(\zeta_{n})-1\right), &  a\in D_3 ,\\
                             -\frac{p-1}{2},&  a\in P, \\
                           \frac{q+1}{2},   &  a\in Q.
                        \end{array}
                        \right.
$$
\end{Lem}
\begin{proof}
 By Lemma \ref{lem2.4}, $aD_0=D_0$ and $aD_1=D_1$ if $a\in D_0$. If $a\in D_0$, then $aP=P$ since $\gcd(a,q)=1$. Hence, we obtain
\begin{align*}
s(\zeta_n^{a})&=1+\sum_{i\in P}\zeta_{n}^{ai}+\sum_{i\in D_0\cup D_1}\zeta_{n}^{ai}\\
&=1+\sum_{i\in P}\zeta_{n}^{i}+\sum_{i\in D_0\cup D_1}\zeta_{n}^{i}\\
&=s(\zeta_n).
\end{align*}

If $a\in D_1$, by Lemma \ref{lem2.4}, $aD_0=D_1$ and $aD_1=D_2$.
By \eqref{eq3.1}, we have
\begin{align*}
s(\zeta_n^{a})&=1+\sum_{i\in P}\zeta_{n}^{ai}+\sum_{i\in D_0\cup D_1}\zeta_{n}^{ai}\\
&=\sum_{i\in D_1}\zeta_{n}^{i}+\sum_{D_2}\zeta_{n}^{i}\\
&=t(\zeta_{n}).
\end{align*}

If $a\in D_2$, by Lemma \ref{lem2.4}, $aD_0=D_2$ and $aD_1=D_3$. By \eqref{eq3.1} and  \eqref{eq3.3}, we get
\begin{align*}
s(\zeta_n^{a})&=1+\sum_{i\in P}\zeta_{n}^{ai}+\sum_{i\in D_0\cup D_1}\zeta_{n}^{ai}\\
&=\sum_{i\in D_2}\zeta_{n}^{i}+\sum_{i\in D_3}\zeta_{n}^{i}\\
&=1-\sum_{i\in D_0\cup D_1}\zeta_{n}^{i}\\
&=1-s(\zeta_{n}).
\end{align*}

If $a\in D_3$, by Lemma \ref{lem2.4}, $aD_0=D_3$ and $aD_1=D_0$. By \eqref{eq3.3}, we have
\begin{align*}
s(\zeta_n^{a})&=1+\sum_{i\in P}\zeta_{n}^{ai}+\sum_{i\in D_0\cup D_1}\zeta_{n}^{ai}\\
&=\sum_{i\in D_3}\zeta_{n}^{i}+\sum_{ D_0}\zeta_{n}^{i}\\
&=1-t(\zeta_{n}).
\end{align*}

If $a\in P$, then $aP=P$. Then by \eqref{eq3.1}, we know
\begin{align*}
s(\zeta_n^{a})&=1+\sum_{i\in P}\zeta_{n}^{ai}+\sum_{i\in D_0\cup D_1}\zeta_{n}^{ai}\\
&=\sum_{i\in D_0\cup D_1}\zeta_{n}^{ai}.
\end{align*}
When $s$ ranges over $\{0,1,\ldots, e-1\}$,  $\left(D_0\cup D_1\right)\bmod{q}$ takes on each element of $\{1,2,\ldots, q-1\}$ exactly $(p-1)/2$ times. It follows from \eqref{eq3.1} that
\begin{align*}
\sum_{i\in D_0\cup D_1}\zeta_{n}^{ai}&=\left(\frac{p-1}{2}\bmod{l}\right)\sum_{i\in P}\zeta_{n}^{i}\\
&=-\frac{p-1}{2}\bmod{l}.
\end{align*}

If $a\in Q$, then $aP=\{0\}$. Then
\begin{align*}
s(\zeta_n^{a})&=1+\sum_{i\in P}\zeta_{n}^{ai}+\sum_{i\in D_0\cup D_1}\zeta_{n}^{ai}\\
&=1+(q-1)+\sum_{i\in D_0\cup D_1}\zeta_{n}^{ai}.
\end{align*}
When $s$ ranges over $\{0,1,\ldots, e-1\}$,  $\left(D_0\cup D_1\right)$ $\bmod{p}$ takes on each element of $\{1,2,\ldots, p-1\}$ exactly $(q-1)/2$ times. It follows from \eqref{eq3.1} that
\begin{align*}
s(\zeta_n^{a})&=1+(q-1)+\sum_{i\in D_0\cup D_1}\zeta_{n}^{ai}\\
&=1+(q-1)+\frac{q-1}{2}\sum_{i\in Q}\zeta_{n}^{i}\\
&=\frac{q+1}{2}\bmod{l}.
\end{align*}
\end{proof}
\begin{Lem} [ \cite{HYW}, Lemma 3.3] \label{lem3.2} Let the notations be the same as before. Then
\begin{enumerate}
\item
$-1\in D_0$ if and only if  $p\not\equiv q\pmod{8}$;
\item
$-1\in D_2$ if and only if $p\equiv q\pmod{8}$.
\end{enumerate}
\end{Lem}
\begin{Lem}[\cite{ALW}, Lemmas 2 and 4]\label{lem3.3}
For each $w\in P\cup Q$,
\begin{align*}
|\{D_{i}\cap (D_{j}+w)\}|=\left\{\begin{array}{ll}
                                   \frac{(p-1)(q-1)}{16} & \textrm{ if $i\neq j$}, \\
                                   \frac{(p-1)(q-5)}{16} & \textrm{ if $i=j$ and $p\mid w$}, \\
                                   \frac{(p-5)(q-1)}{16} & \textrm{ if $i=j$ and $q\mid w$}.
                                 \end{array}
                                 \right.
\end{align*}
\end{Lem}

Define
$$
\Delta_{1}=\frac{p-1}{2}\bmod{l},\qquad  \Delta_{2}=\frac{q+1}{2}\bmod{l}.
$$
\begin{Thm}\label{th3.1}
Let $\frac{n-1}{4}\equiv0\pmod{l}$. Then the linear complexity of the sequence $s$ defined by \eqref{df3.1} is given as follows:
\begin{enumerate}
\item when $p\equiv q\pmod{8}$ and $\frac{b}{2}\equiv0\pmod{l}$, we have
$$
L_{l}(s)=\left\{\begin{array}{cc}
              \frac{pq+p+q-1}{2}-\delta, & \textrm{ if $\Delta_1\neq0$, $\Delta_2\neq0$},\\
              \frac{pq-p+q+1}{2}-\delta, & \textrm{ if $\Delta_1\neq0$, $\Delta_2=0$}, \\
             \frac{pq+p-q+1}{2}-\delta,  & \textrm{ if $\Delta_1=0$, $\Delta_2\neq0$}, \\
              \frac{pq-p-q+1}{2}, & \textrm{ if $\Delta_1=0$, $\Delta_2=0$};
            \end{array}
            \right.
$$
\item when $q\not\equiv p\pmod{8}$ and $\frac{a^{2}+3}{4}\equiv0\pmod{l}$,  we have
$$
L_{l}(s)=\left\{\begin{array}{cc}
              \frac{3pq+p+q-1}{4}-\delta, & \textrm{ if $\Delta_1\neq0$, $\Delta_2\neq0$},\\
              \frac{3pq-3p+q+3}{4}-\delta, & \textrm{ if $\Delta_1\neq0$, $\Delta_2=0$}, \\
             \frac{3pq+p-3q+3}{4}-\delta,  & \textrm{ if $\Delta_1=0$, $\Delta_2\neq0$}, \\
              \frac{3pq-3p-3q+5}{4}, & \textrm{ if $\Delta_1=0$, $\Delta_2=0$};
            \end{array}
            \right.
$$
\item when $p\equiv q\pmod{8}$ and $\frac{b}{2}\not\equiv0\pmod{l}$ or $q\not\equiv p\pmod{8}$ and $\frac{a^{2}+3}{4}\not\equiv0\pmod{l}$, we have
$$
L_{l}(s)=\left\{\begin{array}{cc}
              n-\delta, & \textrm{ if $\Delta_1\neq0$, $\Delta_2\neq0$},\\
             n+1-p-\delta, & \textrm{ if $\Delta_1\neq0$, $\Delta_2=0$}, \\
             n+1-q-\delta,  & \textrm{ if $\Delta_1=0$, $\Delta_2\neq0$}, \\
             n+2-p-q, & \textrm{ if $\Delta_1=0$, $\Delta_2=0$},
            \end{array}
            \right.
$$
\end{enumerate}
where $a$, $b$ are the integers  defined in \eqref{eq3.17} and  $\delta$ is defined in \eqref{eq3.1.1}.
\end{Thm}
\begin{proof}
By definition, we have
$$
s(\zeta_{n})^{2}=\left(\sum_{i\in D_0, j\in D_0}+\sum_{i\in D_0, j\in D_0}+\sum_{i\in D_0, j\in D_0}+\sum_{i\in D_0, j\in D_0}\right)\zeta_{n}^{i+j}.
$$

We first prove prove the conclusions for the case that $q\equiv p\pmod{8}$. In this case, by Lemma \ref{lem3.2}, $-1\in D_0$ and $b$ must be even.  By Lemmas \ref{lem2.2}  and \ref{lem3.3} , we have
$$
\begin{array}{lll}
s(\zeta_{n})^{2}&=&\left(\sum_{i\in D_0, j\in D_0}+\sum_{i\in D_0, j\in D_1}+\sum_{i\in D_1, j\in D_0}+\sum_{i\in D_1, j\in D_1}\right)\zeta_{n}^{i-j}\\
&=&\left(  (0,0)+(3,3)+(0,1)+(3,0)\right)\sum_{i\in D_0}\zeta_{n}^{i}\\
& &+\left( (01,0)+(0,3)+(1,1)+(0,0)\right)\sum_{i\in D_1}\zeta_{n}^{i}\\
& &+\left((2,0)+(1,3)+(2,1)+(1,0)\right)\sum_{i\in D_2}\zeta_{n}^{i}\\
& &+\left(3,0)+(2,3)+(3,1)+(2,0)\right)\sum_{i\in D_3}\zeta_{n}^{i}\\
& &+2\left(\frac{(p-1)(q-1)}{16}\sum_{i\in P}\zeta_{n}^{i}\right)+2\left(\frac{(p-1)(q-1)}{16}\sum_{i\in P}\zeta_{n}^{i}\right)\\
& &+2\left(\frac{(p-1)(q-5)}{16}\sum_{i\in P}\zeta_{n}^{i}\right)+2\left(\frac{(p-5)(q-1)}{16}\sum_{i\in Q}\zeta_{n}^{i}\right)+|D_0|+|D_1|\\
&=& s(\zeta_{n})+b\sum_{i\in D_0\cup D_2}\zeta_{n}^{i}-\frac{b}{2}+M+\frac{p+q-2}{2}.
\end{array}
$$
From $\frac{n-1}{4}\equiv0\pmod{l}$, we know $M+\frac{p+q-2}{2}\equiv0\pmod{l}$. Hence,
$$
s(\zeta_{n})^{2}=s(\zeta_{n})+b\sum_{i\in D_0\cup D_2}\zeta_{n}^{i}-\frac{b}{2}.
$$
Whence,
\begin{equation}\label{eq3.6}
s(\zeta_{n})(s(\zeta_{n})-1)=\frac{b}{2}\left(2\sum_{i\in D_0\cup D_2}\zeta_{n}^{i}-1\right).
\end{equation}
Note that $\frac{n-1}{4}\equiv0\pmod{l}$.
By Lemma \ref{lem3.1}, we have $\sum_{i\in D_0\cup D_2}\zeta_{n}^{i}\in \{1,0\}$. It follows that
\begin{equation}\label{eq3.10}
2\sum_{i\in D_0\cup D_2}\zeta_{n}^{i}-1\in \{1,-1\}.
\end{equation}

Similarly, we have
\begin{equation}\label{eq3.7}
t(\zeta_{n})(t(\zeta_{n})-1)=-\frac{b}{2}\left(2\sum_{i\in D_0\cup D_2}\zeta_{n}^{i}-1\right).
\end{equation}

By Lemma \ref{lem2.3}, \eqref{eq3.6} and \eqref{eq3.7}, we know when $\frac{b}{2}\equiv0\pmod{l}$, there are exactly half of $a's$ with $a\in \mathbb{Z}_{n}^{*}$  such that $s(\zeta_{n}^{a})=0$.
When $\frac{b}{2}\not\equiv0\pmod{l}$, $s(\zeta_{n}^{a})\neq0$ for all $a\in \mathbb{Z}_{n}^{*}$. Then the desirable results on the linear complexity  of the sequence $s$ follow from  \eqref{linearcomp}, \eqref{eq3.16}  and Lemma \ref{lem2.3}.

Now, we prove the conclusions for the case that $q\not\equiv p\pmod{8}$. In this case, by Lemma \ref{lem3.2}, $-1\in D_2$. It follows from Lemmas \ref{lem2.2} and \ref{lem3.3} that
$$
\begin{array}{lll}
 s(\zeta_{n})^{2}&=&\left(\sum_{i\in D_0, j\in D_2}+\sum_{i\in D_0, j\in D_3}+\sum_{i\in D_1, j\in D_2}+\sum_{i\in D_1, j\in D_3}\right)\zeta_{n}^{i-j}\\
   & = &\left((2,2)+(1,1)+(2,3)+(1,2)\right)\sum_{i\in D_0}\zeta_{n}^{i}  \\
   &  &+\left((3,2)+(2,1)+(3,3)+(2,2)\right)\sum_{i\in D_1}\zeta_{n}^{i}  \\
  &  & +\left((0,2)+(3,1)+(0,3)+(3,2)\right)\sum_{i\in D_1}\zeta_{n}^{i} \\
   &  &+\left((1,2)+(0,1)+(1,3)+(0,2)\right)\sum_{i\in D_1}\zeta_{n}^{i}  \\
   &  &+4\left(\frac{(p-1)(q-1)}{16}\sum_{i\in P}\zeta_{n}^{i}\right)+4\left(\frac{(p-1)(q-1)}{16}\sum_{i\in Q}\zeta_{n}^{i}\right) \\
   & = & s(\zeta_{n})+\frac{b\left(2\sum_{i\in D_0\cup D_2}\zeta_{n}^{i}-1\right)-1}{2}+M-\frac{(p-1)(q-1)}{2}.
\end{array}
$$
From $\frac{n-1}{4}\equiv0\pmod{l}$, we have $M-\frac{(p-1)(q-1)}{2}\equiv0\pmod{l}$.
Hence,
$$
s(\zeta_{n})^{2}=s(\zeta_{n})+\frac{b\left(2\sum_{i\in D_0\cup D_2}\zeta_{n}^{i}-1\right)-1}{2}.
$$
Whence,
\begin{equation}\label{eq3.8}
s(\zeta_{n})(s(\zeta_{n})-1)=\frac{b\left(2\sum_{i\in D_0\cup D_2}\zeta_{n}^{i}-1\right)-1}{2}.
\end{equation}
Similarly, we get
\begin{equation}\label{eq3.9}
t(\zeta_{n})(t(\zeta_{n})-1)=-\frac{b\left(2\sum_{i\in D_0\cup D_2}\zeta_{n}^{i}-1\right)+1}{2}.
\end{equation}
Since $\gcd(p-1,q-1)=4$ and $p\equiv q+4\pmod{8}$, we get  $p\equiv1$ or $5\pmod{8}$. Hence, $n=pq\equiv5\pmod{8}$. Together with $(n-1)/4\equiv0\pmod{l}$, we get $l$ is an old prime. By the representation $n=a^{2}+4b^{2}$ with $a\equiv1\pmod{4}$, we have
$$
\frac{n-1}{4}=\frac{a^{2}+3}{4}+(|b|-1)(|b|+1).
$$

Therefore, $\frac{a^{2}+3}{4}\equiv0\pmod{l}$ if and only if $(|b|-1)(|b|+1)\equiv0\pmod{l}$. Since $l$ is odd, we obtain $(|b|-1)(|b|+1)\equiv0\pmod{l}$ if and only if $l$ divides one of  $|b|-1$ and $|b|+1$. By Lemma \ref{lem2.3},  if $\frac{a^{2}+3}{4}\equiv0\pmod{l}$, there are exactly $(p-1)(q-1)/4$  $a$'s such that $s(\zeta_{n}^{a})=0$ for $a\in \mathbb{Z}_{n}^{*}$. If $\frac{a^{2}+3}{4}\not \equiv0\pmod{l}$, $s(\zeta_{n}^{a})\neq0$ for all $a\in \mathbb{Z}_{n}^{*}$.
Then the desirable conclusions on the linear complexity of the sequence $s$ follow from  \eqref{linearcomp}, \eqref{eq3.16}, \eqref{eq3.10} and  Lemma \ref{lem2.3}.
\end{proof}
\begin{Rem}
If  $\frac{n-1}{4}\equiv0\pmod{l}$, by Lemma $3.14$ in \cite{Ding12}, we get $l\in D_0$. For the case $l\notin D_0$, the linear complexity of the sequence $s$ defined by \eqref{df3.1} can also be determined in the following theorem.
\end{Rem}
\begin{Thm}\label{thm3.2}
If $l\notin D_0$, then for the sequence $s$ defined in \eqref{df3.1}, we have
$$
L_{l}(s)=\left\{\begin{array}{ll}
                      n-\delta, & \textrm{ if $\Delta_1\neq0$, $\Delta_2\neq0$},\\
             n+1-p-\delta, & \textrm{ if $\Delta_1\neq0$, $\Delta_2=0$}, \\
             n+1-q-\delta,  & \textrm{ if $\Delta_1=0$, $\Delta_2\neq0$}, \\
             n+2-p-q, & \textrm{ if $\Delta_1=0$, $\Delta_2=0$},
                  \end{array}
                  \right.
$$
\end{Thm}
\begin{proof}
If $l\notin D_0$, then there exists $i\in\{1,2,3\}$ such that $i\in D_{i}$. No matter what $i$ is, there exists $j\in \{1,2,3\}$ satisfying the congruence equation $ij\equiv2\pmod{4}$. Then $l^{j}\in D_2$ and
$$
s^{l^{j}}(\zeta_{n})=s(\zeta_{n}^{l^{j}})=1-s(\zeta_{n}).
$$
This implies $s(\zeta_{n})\notin \{0,1\}$.

Similarly, $t^{l^{j}}(\zeta_{n})=t(\zeta_{n}^{l^{j}})=1-t(\zeta_{n})$ and $t(\zeta_{n})\notin \{0,1\}$. Hence,  $s(\zeta_{n})\neq0$ for all $a\in\mathbb{Z}_{n}^{*}$. Combining  \eqref{linearcomp} and Lemma \ref{lem2.3}, we have the desirable results.
\end{proof}

Define
\begin{eqnarray*}
d_{j}(x)=\prod_{i\in D_{j}}(x-\zeta_{n}^{i})
\end{eqnarray*}
for $j\in\{0,1,2,3\}$. If $l\in D_0$,  it is easily proved that $d_{j}(x)\in$ GF($l^{m}$)[x] for all $j$.

Let $d(x)=\prod_{j=0}^{3}d_{j}(x)$, then $d(x)\in$ GF($l^{m}$)[x]. We have then
$$
x^{n}-1=\prod_{i=0}^{n-1}(x-\zeta_{n}^{i})=\frac{(x^{q}-1)(x^{p}-1)d(x)}{x-1}.
$$

\begin{Lem}[\cite{HYW}, Lemma 3.3]\label{lem3.4}
Let notations be the same as before. Then
\begin{enumerate}
\item $2\in D_0\cup D_2$ if and only if $p\equiv q\pmod{8}$;
\item $2\in D_1\cup D_3$ if and only if $p\not\equiv q\pmod{8}$.
\end{enumerate}
\end{Lem}
\begin{Lem}[\cite{HYW}, Lemma 3.5]
Let $p\equiv q\pmod{8}$. Then there are exactly two representations over $\mathbb{Z}$
$$
pq=a^{2}+4b^{2}=a'^{2}+4b'^{2}, \qquad a\equiv a^{'}\equiv1\pmod{4},
$$
where one of $b$ and $b^{'}$ is divided by $4$ and another is exactly divided by $2$.

\end{Lem}
\begin{Lem}[\cite{HYW}, Corollary 3.9]\label{lem3.5}
Let $p\equiv q\pmod{8}$. Fix a common primitive root  $g$ of $p$ and $q$. Then $2\in D_0$ if and only if the generalized cyclotomic numbers in  Lemma $\ref{lem2.2}$ depend on the decomposition $n=a^{2}+4b^{2}$ with $4|b$; $2\in D_2$, if and only if the generalized cyclotomic numbers depend on the decomposition $n=a^{2}+4b^{2}$ with $2||b$.
\end{Lem}

After the preparations above, we are ready to compute the linear complexity and minimal polynomials of the sequence defined in \eqref{df3.1} over $\mathbb{F}_2$.

\begin{Cor}
Let $l=2$ and $n\equiv1\pmod{8}$. Then $p\equiv q\pmod{8}$ and  the linear complexity and minimal polynomials of the sequence $s$ defined by \eqref{df3.1} are given as follows:
\begin{enumerate}
\item when  $2\in D_0$, then
$$
L_{2}(s)=\frac{pq-q+p+1}{2},
$$
$$
m(x)=\left\{\begin{array}{ll}
              \frac{(x^{n}-1)(x-1)}{(x^{q}-1)d_2(x)d_1(x)}, & \textrm{ if $s(\zeta_{n})=1$ and $t(\zeta_{n})=0$}, \\
               \frac{(x^{n}-1)(x-1)}{(x^{q}-1)d_0(x)d_3(x)}, & \textrm{ if $s(\zeta_{n})=0$ and  $t(\zeta_{n})=1$},\\
               \frac{(x^{n}-1)(x-1)}{(x^{q}-1)d_0(x)d_1(x)}, & \textrm{ if $s(\zeta_{n})=0$ and  $t(\zeta_{n})=0$},\\
               \frac{(x^{n}-1)(x-1)}{(x^{q}-1)d_2(x)d_3(x)}, & \textrm{ if $s(\zeta_{n})=1$ and  $t(\zeta_{n})=1$};
            \end{array}
            \right.
$$
\item when  $2\in D_2$ ,  then
$$
 L_{2}(s)=n+1-q,\ m(x)=\frac{(x^{n}-1)(x-1)}{x^{q}-1}.
$$

\end{enumerate}
\end{Cor}
\begin{proof}
If $p\not\equiv q\pmod{8}$, then $p\equiv q+4\pmod{8}$ and  $n=pq\equiv 5\pmod{8}$. This contradicts with $n\equiv1\pmod{8}$. Hence, $p\equiv q\pmod{8}$. And in this case, by Lemma \ref{lem3.4}, $2\in D_0\cup D_2.$

By \eqref{eq3.6}, \eqref{eq3.7} and Lemma \ref{lem3.5}, if $2\in D_0$, then
$$
s(\zeta_n)\in \{0,1\}, \ \ t(\zeta_n)\in \{0,1\},
$$
and if $2\in D_2$, we obtain
$$
s(\zeta_n)\notin \{0,1\} \ \ t(\zeta_n)\notin \{0,1\}.
$$
Then the desired conclusions on the linear complexity and the minimal polynomial of the sequence $s^{\infty}$ follow from  \eqref{linearcomp} and Lemma \ref{lem2.3}.
\end{proof}

If $p\not\equiv q\pmod{8}$, then $n=pq\equiv5\pmod{8}$ and  by Lemma $2\in D_1 \cup D_3$. Hence, by Theorem \ref{thm3.2}, we have the following corollary.
\begin{Cor}
Let $l=2$ and $n\equiv5\pmod{8}$. Then   for the sequence $s$ defined by \eqref{df3.1}, we have
$$
L_{2}(s)=n+1-q, \qquad m(x)=\frac{(x^{n}-1)(x-1)}{x^{q}-1}.
$$
\end{Cor}
\begin{Exp}
Let $(p,q)=(5,13)$ and $(g_1,g_2)=(2,2)$. Then $a=1$ and $b=4$. If $l=2$,  Magma program  shows that $2\in D_0$ and $L_2(s)=29$. If $l=3$, then $b/2\not\equiv0\bmod{l}$ and $L_3(s)=65$.
\end{Exp}
\begin{Exp}
Let $(p,q)=(5,17)$ and $(g_1,g_2)=(2,3)$. Then $a=-7$ and $b=3$. If $l=2$, it can be easily checked that $2\in D_3$ and $L_2(s)=69$. If $l=7$, then $(a^{2}+3)/4\not\equiv0\bmod{l}$ and $L_7(s)=85$.
\end{Exp}
\section{Concluding Remarks}

In this paper, we determined the linear complexity of generalized cyclotomic sequences of order $4$ over $\mathbb{F}_{l}$ under certain conditions. Results show that these sequences have high linear complexity for a large part of prime numbers $p$ and $q$. Recently, periodic sequences were used to construct cyclic codes and the dimension of the cyclic code is closely related to the linear complexity of the corresponding sequence over a finite field \cite{Ding12,Ding13}. This paper could be viewed as an application of this idea.


\end{document}